\documentstyle[preprint,pre,aps,eqsecnum,epsf,graphicx,color]{revtex}
\tightenlines
\newcommand{\beq}{\begin{eqnarray}}
\newcommand{\eeq}{\end{eqnarray}}

\newcommand{\pardis}{\langle \mu \rangle}

\newcommand{\ie}{{\it i.e.\ }}

\newcommand{\real}{{\sf I}\kern-.12em{\sf R}}
\newcommand{\comp}{{\sf I}\kern-.50em{\sf C}}
\newcommand{\unity}{{\sf I}\kern-.54em{\sf 1}}
\def\spose#1{\hbox to 0pt{#1\hss}}
\def\ltapprox{\mathrel{\spose{\lower 3pt\hbox{$\mathchar"218$}}
\raise 2.0pt\hbox{$\mathchar"13C$}}}

\begin{document}

\rightline{GEF-TH 12/06}
\rightline{UB-ECM-PF 06/23}
\centerline{\bf Dual superconductivity and vacuum properties in
Yang--Mills theories}
\vskip 5mm
\centerline{A. D'Alessandro$^{a}$, M. D'Elia$^{a}$
and L. Tagliacozzo$^{b}\,\,$\footnote{E-mail addresses:
  adales@ge.infn.it, delia@ge.infn.it, luca@ecm.ub.es}} 
\centerline{\it $^a$Dipartimento di Fisica, Universit\`a di Genova and
  INFN, Sezione di Genova,}
\centerline{\it Via Dodecaneso 33, I-16146 Genova, Italy}
\centerline{\it $^b$ Departament d'Estructura i Constituents de la 
Mat\`eria, Universitat de Barcelona,}
\centerline{\it 647, Diagonal, 08028, Barcelona, Spain.}

\begin{abstract}
We address, within the dual superconductivity model for color
confinement, the question whether the Yang-Mills vacuum behaves as a
superconductor of type I or type II. 
In order to do that we compare, for the theory with gauge group
$SU(2)$, the determination of the field penetration
depth $\lambda$ with that of the superconductor correlation length
$\xi$. The latter is obtained by measuring the temporal correlator 
of a disorder parameter developed by the Pisa group
to detect dual superconductivity.
The comparison places the  vacuum close to the border
between type I and type II and marginally on the type II side. We 
also  check our results against the study of 
directly measurable effects such as the interaction between
two parallel flux tubes, obtaining consistent indications 
for a weak repulsive behaviour.
Future strategies to improve our investigation are discussed.
\end{abstract}

\section{Introduction}

Color confinement emerges as a fundamental property of strongly
interacting matter from experimental facts, like for instance the absence of
fractionally charged particles. Even if lattice simulations provide
evidence that confinement is realized in the theory of strong interactions, a
full theoretical explanation of it starting from QCD first principles is
still lacking. However models exist which relate confinement to some
property of the fundamental state of the theory. One of those models
is based on dual superconductivity of the QCD
vacuum~\cite{thooft75,mandelstam,parisi}:
according to this model confinement of color is due to the
spontaneous breaking of a magnetic simmetry which yields a
nonvanishing magnetically charged Higgs condensate. The dual Meissner
effect compels the electric field between static colored charges in
narrow flux tubes, giving rise to a linearly rising potential and to
confinement. The broken magnetic group is chosen by a procedure
known as {\it Abelian projection}~\cite{thooft81}: a local operator
$\phi (x)$ transforming in the adjoint representation is
diagonalized, leaving a residual $U(1)^{N_c-1}$ gauge symmetry.

A superconductor is characterized by two fundamental parameters,
the correlation length $\xi$ of the Higgs condensate and the field
penetration depth $\lambda$: they determine whether the
superconductor is of type I ($\xi > \lambda$) or type II ($\xi <
\lambda$). In a superconductor of type I an external field $B$ is
always expelled from the medium  till a critical
value $B_c$ beyond which superconductivity disappears. In a superconductor
of type II there are instead two different critical values $B_{c1}$
and $B_{c2}$, and for $B_{c1} < B < B_{c2}$ the external field
can penetrate the medium in the form of Abrikosov flux tubes, without
disrupting superconductivity. Another relevant property of type II
superconductors is the repulsive interaction between two parallel
flux tubes, which is instead attractive for type I superconductors.

In the framework of the dual superconductor model, understanding
whether the QCD vacuum behaves as a type I or a type II
superconductor is an issue which can help clarifying the dynamics of
color confinement and of flux tube interactions. The question can
in principle be answered by QCD numerical lattice simulations and
several efforts have been done in the past in that direction, mostly
for the pure gauge theory with 2 colors. A direct way to determine
$\lambda$ is a lattice
analysis~\cite{suz94,cea95,bali98,suz99,suz03,hay05} of the flux
tube which is formed between two static color charges: the
longitudinal (chromo)electric field $E_z$ decays asymptotically as
$E_z=A K_0(\frac{d}{\lambda})$ at a radial distance $d$ from the
tube axis.

The determination of $\xi$ is less straightforward:
 this parameter has been found in literature mostly either through an
analysis of violations of the $E_z=A K_0(\frac{d}{\lambda})$
behavior close to the center of the flux tube~\cite{cea95,hay05} or
through some global fit to the whole set of Ginzburg-Landau
equations~\cite{bali98,suz99,suz03}; a determination based on a
direct analysis of the condensate distribution around the flux tube
has also appeared recently~\cite{suz05}.  An approximate picture has
emerged placing the $SU(2)$ Yang-Mills vacuum roughly at the boundary
between a type I and a type II dual superconductor.

In the present study we consider the case of $SU(2)$ pure gauge theory
and follow a different strategy, aimed at
determining the mass of the Higgs field $m_H  = 1/\xi$ through
the analysis of the temporal correlator of an observable directly
coupled to it: that is the  operator $\mu$ developed by the
Pisa group which creates a magnetic monopole (see Ref.~\cite{dig97}
for a detailed discussion about its definition and also 
Ref.~\cite{Frohlich:2000zp} and~\cite{CC} for related parameters).
Its vacuum expectation value ({\em v.e.v.}) $\langle \mu \rangle$
is a good disorder parameter detecting
dual superconductivity ($\langle \mu \rangle \neq 0$) and
the transition to the deconfined - normal conducting phase
($\langle \mu \rangle = 0$) both in pure gauge
theory~\cite{artsu2,artsu3,artran} and in full
QCD~\cite{artfull,artfull2}.
We will compare results obtained for $\xi$ in this way with those obtained for
$\lambda$ through the usual analysis of the field inside the flux tube.
As a further independent method to characterize the QCD
vacuum, we will also directly study the interaction between flux tubes by
measuring the electric field in presence of two couples of static
charges. Preliminary results concerning the determination of $\xi$
have been reported in Ref.~\cite{qcd05}.

In Section \ref{xi} we will review the definition of the disorder
parameter $\pardis$ and present a determination of $\xi$ based on the
measurement of its temporal correlator. The results obtained for $\xi$
will be compared
in Section \ref{lambda} with those obtained for $\lambda$. Conclusions
concerning the typology of the vacuum will then be checked against
a direct analysis of flux tube interactions in Section \ref{2tube}.
Finally in Section \ref{final} we will present our conclusions
and discuss possible improvements as well as possible
future extensions of our study.

\section{Disorder parameter for dual superconductivity 
and determination of the correlation length $\xi$}
\label{xi}

\subsection{The disorder parameter and its temporal correlator}

A disorder parameter detecting dual superconductivity can be
constructed in terms of an operator $\mu$ which creates a magnetic
charge. It can be defined in the continuum as~\cite{dig97}:
\begin{equation}
\mu^a (\vec x, t) =
 \exp \left( \,\,\, i \int\!\! d\vec y \,\,{\rm Tr} \{ \phi^a(\vec y, t) \vec
 E (\vec y, t)\} \vec b_\perp (\vec y - \vec x) \right) \label{MUA} \,
\end{equation}
where $\phi^a(\vec y, t)$ is the adjoint field defining the abelian
projection,  $ \vec b_\perp $ is the field of a monopole sitting at
$\vec x$ and $\vec E (\vec y, t)$ is the chromoelectric field. 
The construction of $\mu$ is analogous to that of a
translational operator in quantum mechanics: it creates a magnetic
monopole by  shifting the quantum vector potential field by
the classical field $ \vec b_\perp $.
On the lattice correlation functions of $\mu (\vec x, t)$ can be
written as (see \cite{dig97,artsu2,artsu3,artran} for details):
\beq
\langle \bar{\mu}(t',\vec{x}')\mu(t,\vec{x}) \rangle = \frac{\tilde Z}{Z}
= \frac{\int \left( {\cal D}U \right)  e^{-\beta {\tilde S}}}{\int \left( {\cal D}U \right)  e^{-\beta {S}}}
\label{zratio}
\eeq
where $S$ is the usual pure gauge action and $\tilde S$ differs from
$S$ only at time slices $t$ and $t'$. In particular, in the abelian
projected gauge the temporal plaquettes
\begin{eqnarray}
\Pi_{i0} (\vec{y},y_0)
= U_i (\vec{y},y_0) U_0 (\vec{y} + \hat{\imath},y_0)
U_i^\dagger(\vec{y},y_0 + \hat{0}) U_0^\dagger (\vec{y},y_0)
\end{eqnarray}
are changed by substituting
\begin{eqnarray}
U_i(\vec{y},y_0) \to \tilde{U}_i(\vec{y},y_0) \equiv
U_i(\vec{y},y_0) e^{i T b^i_\perp (\vec{y} - \vec{x})}
\label{modlink}
\end{eqnarray}
where $T$ is the diagonal gauge group
generator corresponding to the monopole species chosen
($T_3 = \lambda_3/2$ is the only possible choice for the $SU(2)$ gauge group) and
$b^i_\perp$ is the transverse vector field corresponding
to the monopole (antimonopole) sitting at $t$ ($t'$) and $\vec x$.

The numerical study of the temporal correlator of $\mu$ as a mean to
determine the monopole mass has already been considered 
for the $U(1)$ pure gauge theory in 4 dimensions~\cite{dig97,luca}.
In the confined phase, where dual superconductivity is at work,
$\pardis \neq 0$. Therefore at large temporal distances the correlator
$\langle \bar{\mu}(t,\vec{x})\mu(0,\vec{x}) \rangle$ is dominated, by
cluster property, by a term $\pardis^2$ plus a function which vanishes
exponentially according to the mass $M$ of the
lightest state coupled to $\mu$.
Taking into account that we are computing
a point-point correlator 
instead of a zero momentum one 
and neglecting the possible presence of excited states, 
we will consider as the simplest possible ans\"atz the leading
large distance behaviour of the two point correlation function:
\begin{eqnarray}
\langle \bar{\mu}(t,\vec{x})\mu(0,\vec{x}) \rangle \simeq
  \langle \mu \rangle^2 + \gamma \frac{e^{-M t}}{t^{3/2}} \, .
\label{betcorrel}
\end{eqnarray}

Since the ratio of partition
functions in Eq.~(\ref{zratio}) is an exponentially noisy quantity,
it is not easy to measure the correlator
$\langle \bar\mu \mu \rangle$ directly and one usually measures:
\beq
\label{defrho}
\rho = \frac{d}{d \beta} \ln \langle \bar\mu \mu \rangle =
 \langle S \rangle_S -
\langle \tilde{S} \rangle_{\tilde{S}} \,
\eeq
where the subscript indicates the action that is used in the Boltzmann weight.
The behaviour expected for $\rho$ at large $t$ can be easily
derived from Eq.~(\ref{betcorrel}); after introducing the adimensional
lattice quantities $\hat M = a M$, $\hat t = t / a$,
$\vec n = \vec x /a$ and after rescaling $\gamma \to a^{3\over2}\gamma$,
where $a$ is the lattice spacing, one obtains:
\beq
\label{rhot}
\rho ( \hat{t} ) \equiv \frac{d}{d \beta}
\ln \langle \bar{\mu}(\hat t,\vec{n})\mu(0,\vec{n}) \rangle \simeq
\frac{A + B \, e^{-\hat M \hat t}/~\hat t^{1/2} + C \,  {e^{-
      \hat M \hat t}}/~{\hat t^{3/2}}}
     {\langle \mu \rangle^2 + \gamma \, {e^{- \hat M \hat t}}/~{{\hat t}^{3/2}}}
\eeq
where
\beq
A =  \frac{d \langle \mu \rangle^2}{d \beta}\,;
\,\,\,\,\, B = - \gamma \frac{d \hat{M}}{d \beta} =
- \gamma M \frac{d a}{d \beta}\,;
\,\,\,\,\,
C = \frac{d \gamma}{d \beta}\,.
\eeq
Eq.~(\ref{rhot}) will be the basis for our fits to the temporal
correlator $\rho(\hat t)$, which will be discussed in Section~\ref{rhoresults}.
Results obtained through a different observable, also related to the
temporal correlator in Eq.~(\ref{betcorrel}) and introduced in
Ref.~\cite{luca}, will be presented and discussed in
Section~\ref{rhotilde}.

As a result of our fits we will obtain an estimate of
$\xi_\mu \equiv a \hat M^{-1}$. The fact that $\langle \mu \rangle$ is a
good disorder parameter for dual superconductivity means that it is
surely coupled to the condensing Higgs field. The natural expectation is
therefore that $\xi_\mu = \xi$, which is true apart from the
unlikely case where the actual field which condenses in
the vacuum does not coincide with the lowest mass state having the
same quantum numbers (in that case one would have  $\xi < \xi_\mu$).

\subsection{Monte Carlo simulations and discussion of results}
\label{rhoresults}

We have measured the correlator $\rho(\hat t)$ using a magnetic charge
defined in the so-called random abelian projection, which was proposed in
Ref.~\cite{artran} and is a sort of average over all
possible abelian projections: in that case
one thus does not need to perform any gauge fixing at all, with a great benefit
in computational cost. The dependence of our results on the abelian
projection chosen will be discussed in Section~\ref{fixedgauge}, where
we will make a comparison with results obtained by taking the abelian
projection in the gauge where the Polyakov loop is diagonal.

The correlator $\rho(\hat t)$ is composed of two terms
(see Eq.~(\ref{defrho})):
\beq
\rho(\hat t) =  \langle S \rangle_S -
\langle \tilde{S}(\hat t) \rangle_{\tilde{S}(\hat t)} \, ;
\eeq
since the first term is independent of $\hat t$, we have
only determined the expectation value of the modified action
$\langle \tilde{S}(\hat t) \rangle_{\tilde{S}(\hat t)}$ : 
we notice that a different Monte Carlo simulation is
required for each  value of $\hat t$.

We have performed simulations at four different values of the inverse
bare coupling, $\beta = 2.4, \,2.5115,\, 2.6,\, 2.7$, in
order to eventually check the correct scaling of our results to the
continuum limit.
For the determination of the physical scale we make reference
to the non-perturbative computation of the $\beta-$function and 
to the determination of $T_c/\sqrt{\sigma}$ reported
in Refs.~\cite{fingberg,karsch}, from which we have inferred the following
values of the lattice spacing: $a(\beta = 2.4) \simeq 0.118 \, {\rm fm}$,
$a(\beta = 2.5115) \simeq 0.083 \, {\rm fm}$,
$a(\beta = 2.6) \simeq 0.062 \, {\rm fm}$ and 
$a(\beta = 2.7) \simeq 0.046 \, {\rm fm}$.
 The lattice volumes $N_s^3
\times N_t$ have been chosen so as to have approximately equal spatial
volumes at the three lowest coupling values:
$12^3 \times 16$ at $\beta = 2.4$,
$16^3 \times 20$ at $\beta = 2.5115$
$20^3 \times 20$ at $\beta = 2.6$. At $\beta = 2.7$ we have been
compelled by computational constraints to use again a
$20^3 \times 20$ lattice, which is smaller in physical units,
but is however comparable, as for the spatial size, 
to a $12^3 \times 20$ lattice at $\beta = 2.5115$, where we have
checked that finite size effects in the determination of $\xi$ are
negligible, at least within our statistical uncertainties.
Different values of the magnetic charge $Q$ carried by the monopole
have been used in some cases, in order to check that our results are 
independent of this quantity.

\begin{figure}[htp]
\begin{center}
\vspace{0.5cm}
\begin{tabular}{cc}
\includegraphics[width=7.5cm]{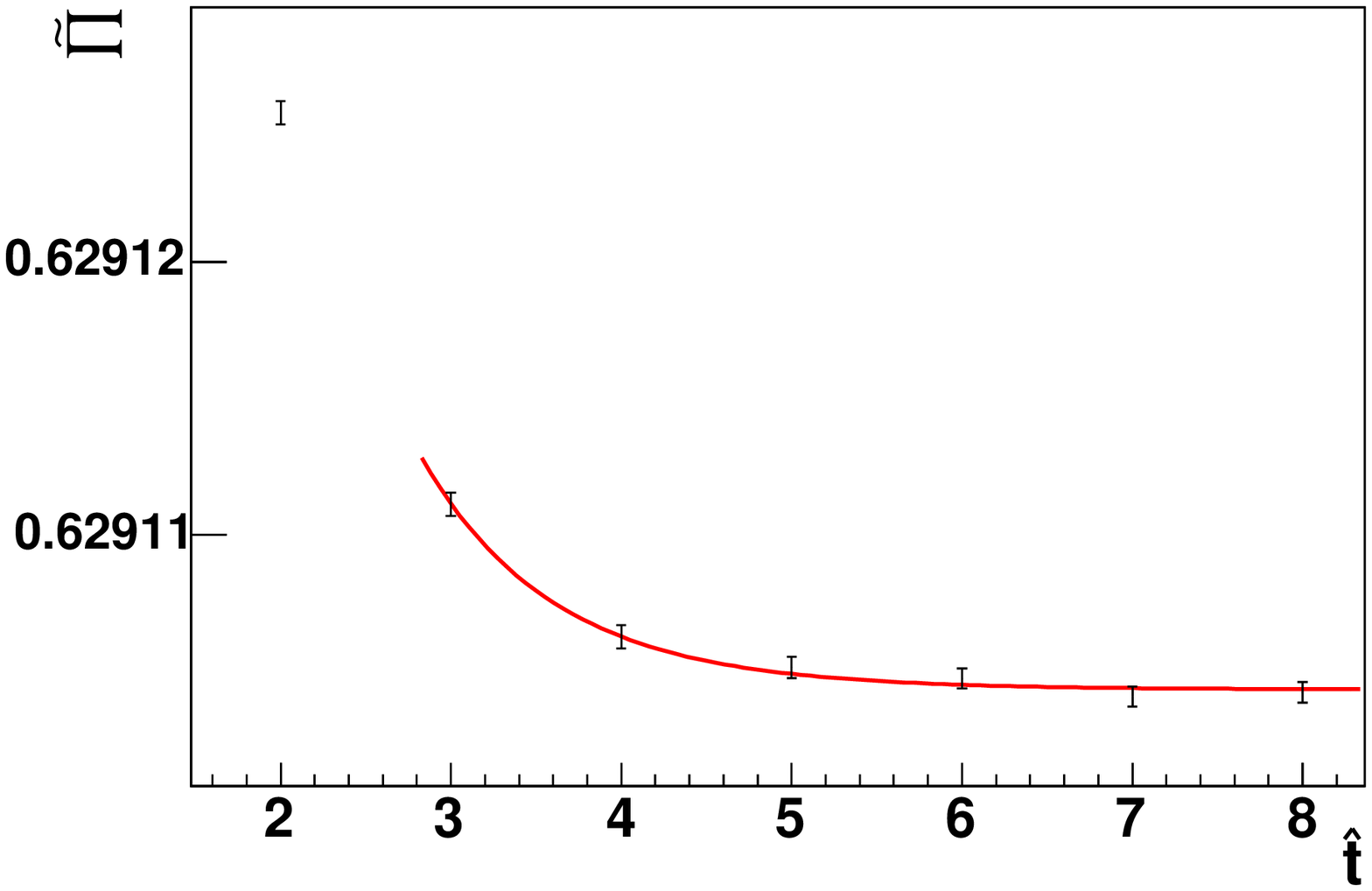}     &  \includegraphics[width=7.5cm]{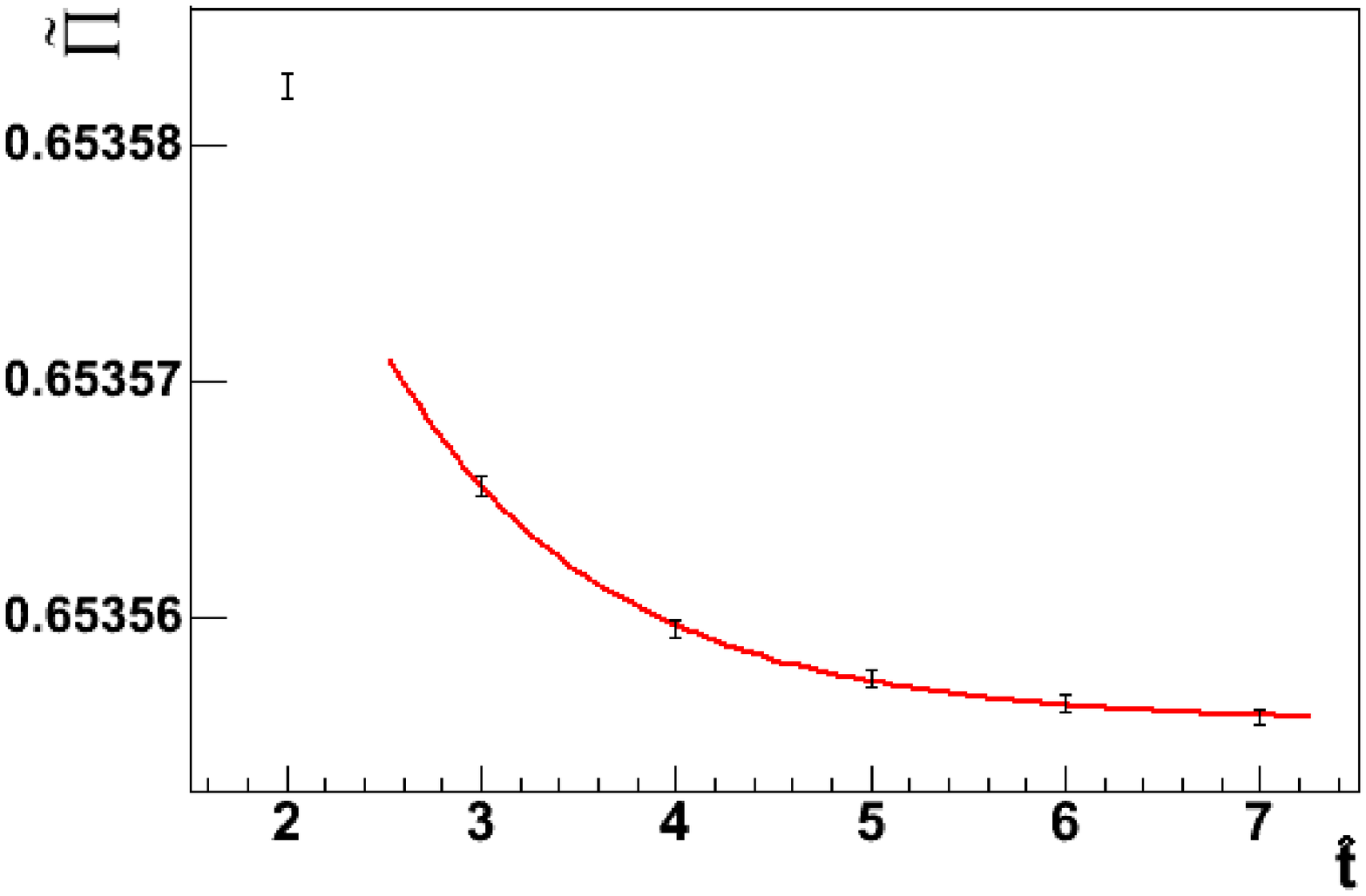} \\
\includegraphics[width=7.5cm]{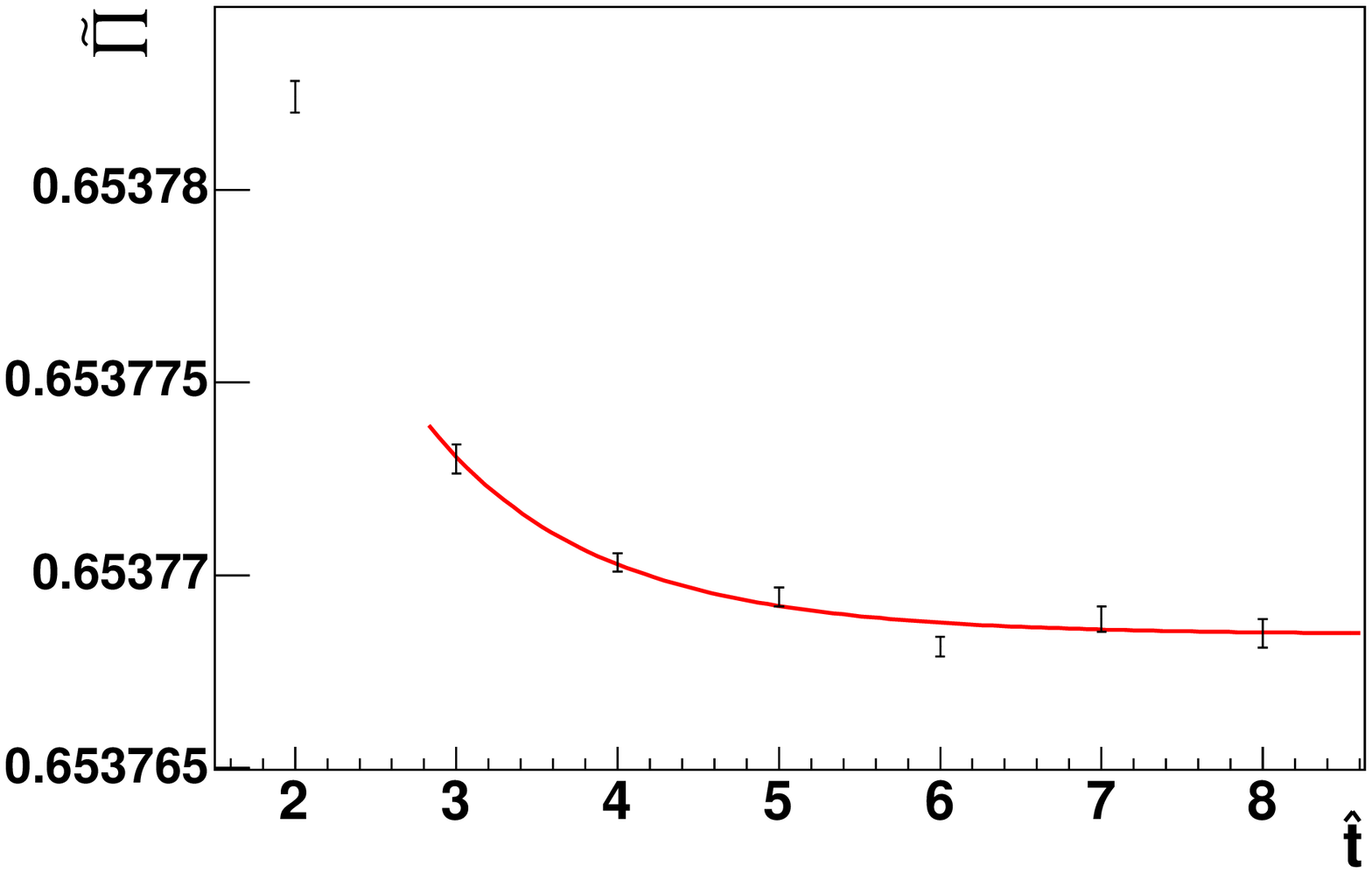}  &
\includegraphics[width=7.5cm]{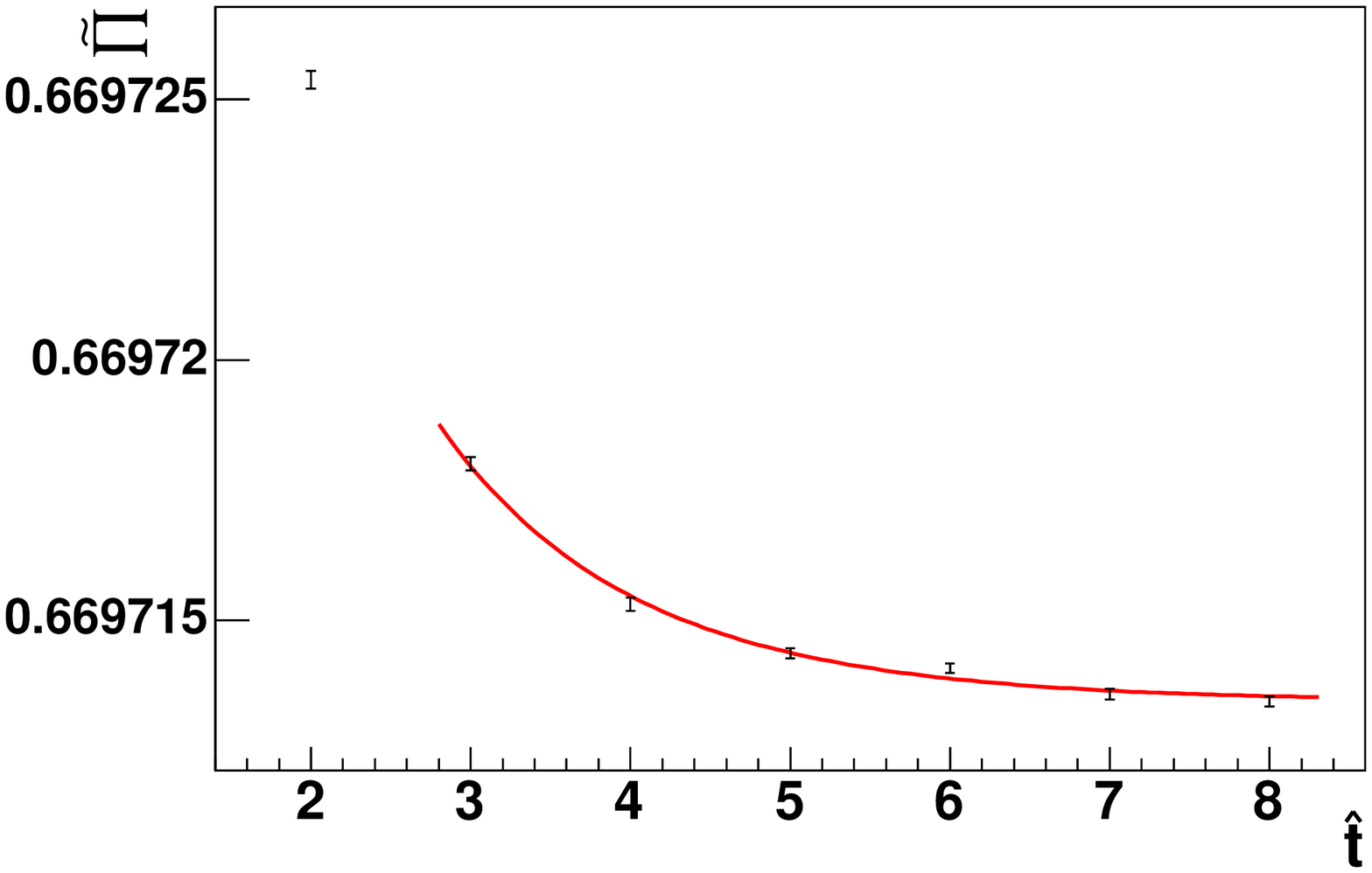} \\
\includegraphics[width=7.5cm]{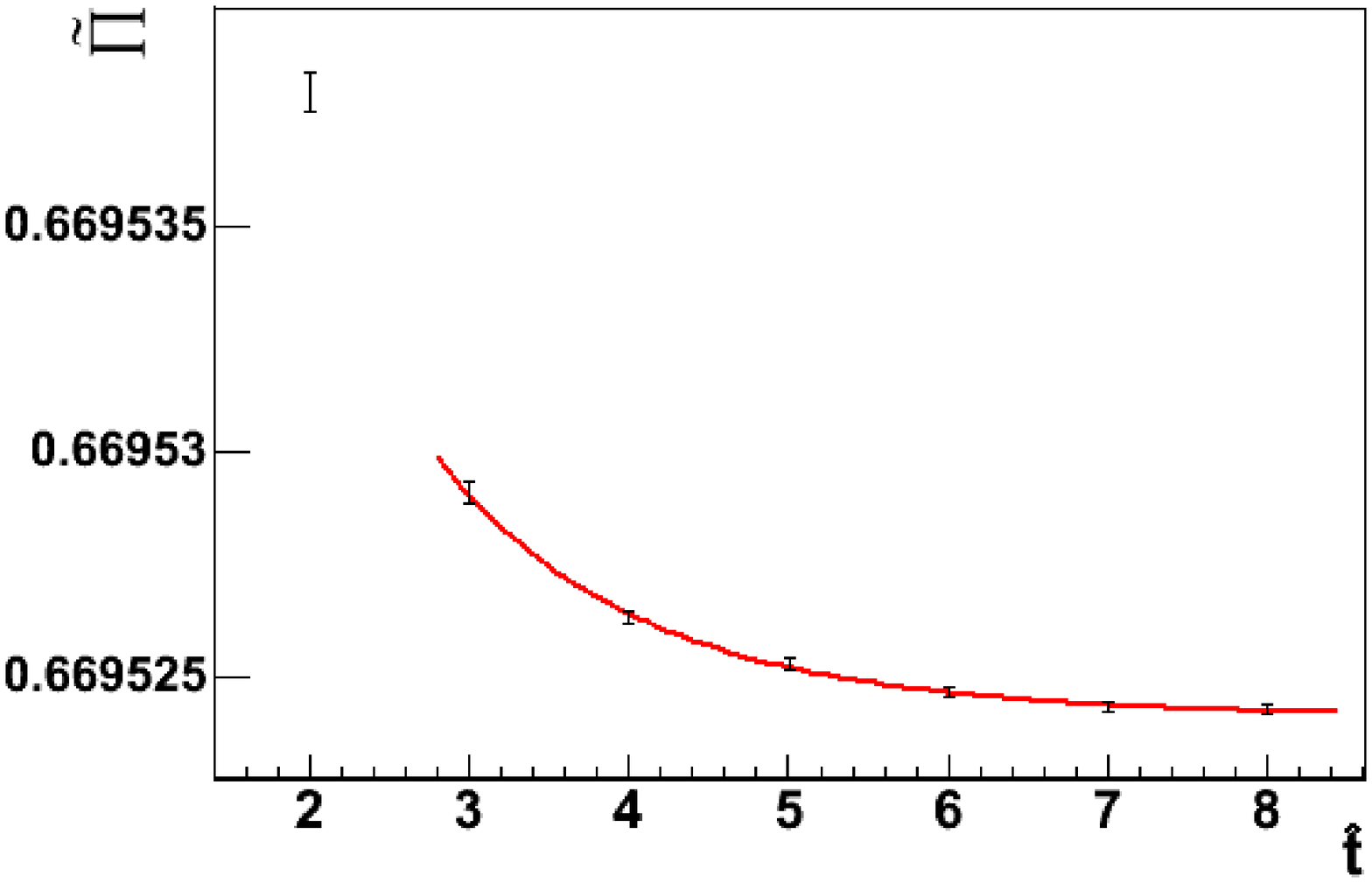}  &
\includegraphics[width=7.5cm]{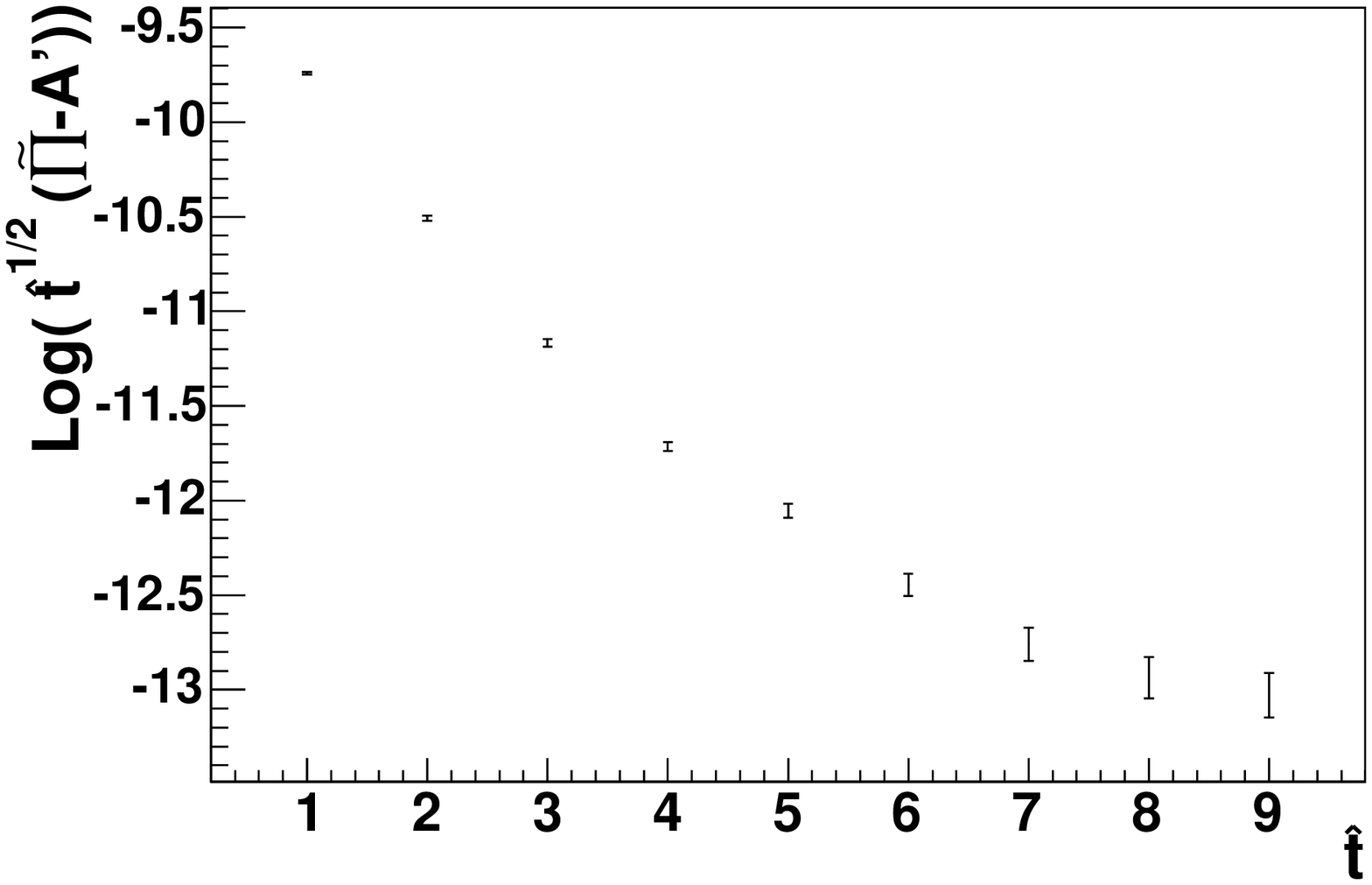} \\
\end{tabular}
\vspace{0.5cm}
\caption{Modified average plaquette as a function of $\hat t$ for a
$12^3\times16$ lattice at $\beta=2.4$ (top left), a
$12^3\times20$ lattice at $\beta=2.5115$ (top right), a
$16^3\times20$ lattice at $\beta=2.5115$ (middle left), a
$20^3\times20$ lattice at $\beta=2.6$ with $Q = 2$ (middle right),
a $20^3\times20$ lattice at $\beta=2.6$ with $Q = 8$ (bottom left) and
a $20^3\times20$ lattice at $\beta=2.7$ with $Q = 8$ (bottom right).
In the last case ($\beta = 2.7$) we have actually plotted the quantity 
$\log (\hat{t}^{1/2} (\tilde\Pi - A`))$ (see Eq.~(\ref{fitfunc})) in
order to highlight the physical signal in the correlator.
\label{xigraph}
}
\end{center}
\end{figure}

The signal obtained for $\langle \tilde{S}(\hat t)
\rangle_{\tilde{S}(\hat t)}$ is mostly made up of a constant background:
it is therefore essential
to reduce the noise as much as possible to obtain a good definition
of the exponentially decaying signal. In order to do that we have
integrated analytically over the probability measure of each gauge
link (as the other links were left fixed), thus obtaining an
improved estimate for the local action density. The typical number
of measurements taken for each  determination of the temporal
correlator has ranged from $10^6$ to about $5 \cdot 10^6$. We report
in Fig.~\ref{xigraph} a summary of the results obtained for the
modified action density $\tilde \Pi (\hat t) \equiv \langle
\tilde{S}(\hat t) \rangle_{\tilde{S}(\hat t)} / 6 V $.

The expected behaviour for $\tilde \Pi (\hat t)$  stems from
Eq.~(\ref{rhot}) by simply adding a constant term; as a matter of
fact, due to the high number of parameters in Eq.~(\ref{rhot}) and
to the poor quality of our signal, we have been able to fit only the
leading large $\hat t$ behaviour of
Eq.~(\ref{rhot}),\footnote{A fit which takes into account also the
  next
to leading term, $e^{-\hat t/\hat \xi}/\hat t^{3/2}$, gives
compatible results, but errors are of the same order of
the fitted values.} which taking into account the periodic boundary
conditions in the time direction is \beq \tilde \Pi (\hat t) = A' + B'
\left(\frac{e^{-\hat t/\hat \xi}}{\hat t^{1/2}} +
\frac{e^{-(N_t-\hat t)/\hat \xi}}{{(N_t-\hat t)}^{1/2}} \right)
\label{fitfunc} \eeq

In Table~\ref{xitable} we report the fit results obtained for $\hat
\xi$ according to Eq.~(\ref{fitfunc}) as a function of the initial
fitting point $\hat t_0$: since we are taking into account only the
leading large $\hat t$ behaviour, and also in order to avoid 
contaminations from higher excited states coupled to $\mu$,
we must search for a plateau in $\hat \xi$ as a function of 
$\hat t_0$. That is not an easy task since, as we will soon discuss,
the correlation length $\xi$ comes out to be of the order of $0.1$ fm: that,
combined with the very low signal/noise ratio characterizing our observable,
leads to a signal which disappears after a few lattice spacings, so
that fits for  $\hat t_0 \ge 4$ are hardly feasible, except for the
highest values of $\beta$ and $Q$ where $\hat \xi$ is larger and the 
signal sharper. 

As a general rule, we have considered for our
determination of $\hat \xi$ the value of $\hat t_0$ after which the
signal does not change considerably within errors:
that corresponds to $\hat t_0 = 2$ for $\beta = 2.4$, 
$\hat t_0 = 3$ for $\beta = 2.5115$ and  $\beta = 2.6$,  
$\hat t_0 = 4$ for $\beta = 2.7$. The corresponding values obtained
for the $\tilde \chi^2$ test are the following:
$\chi^2/\mbox{d.o.f.}(12^3\times16,\beta=2.4)=0.46/3$,
$\chi^2/\mbox{d.o.f.}(12^3\times20,\beta=2.5115)=1.2/2$,
$\chi^2/\mbox{d.o.f.}(16^3\times20,\beta=2.5115)=2.1/5$,
$\chi^2/\mbox{d.o.f.}(20^3\times20,\beta=2.6, Q=2)=2.7/3$,
$\chi^2/\mbox{d.o.f.}(20^3\times20,\beta=2.6, Q=8)=1.3/3$
and $\chi^2/\mbox{d.o.f.}(20^3\times20,\beta=2.7)=6.4/4$.
The two determinations obtained at
$\beta = 2.5115$ on the two different lattice sizes are in
agreement, thus indicating that finite size effects are not important
within our present statistical errors.
The two determinations obtained at $\beta = 2.6$ for two different values
of the monopole charge are nicely compatible within errors, thus
showing no significant dependence of $\xi$ on $Q$.
 
For the two lowest values of $\beta$ the resulting value of $\xi$
is quite close, actually compatible, with the value of the lattice
spacing itself, so that lattice artifacts could play an important
role. The situation improves for $\beta = 2.6$ and $\beta = 2.7$,
which can then be considered as more reliable determinations: we
notice that the two values $\hat t_0 = 3$ and $\hat t_0 = 4$ used
respectively for the determinations at $\beta = 2.6$ and $\beta = 2.7$
are approximately equal when converted in physical units.

In Table~\ref{xiscalingtab} we report a summary of the values
in physical units obtained for $\xi$ as a function of $\beta$,
together with the lattice spacing $a(\beta)$. Our results 
seem compatible, within errors, with the correct scaling to the 
continuum limit; however, taking into account that the 
determinations at the two lowest values of $\beta$ are very close
to the ultraviolet cutoff and that in general our statistical 
uncertainties are still large, a reliable extrapolation
to the continuum limit is still not possible. Rather we give as 
our best determination of the correlation length $\xi$ 
that obtained at $\beta = 2.7$, which, using a conservative 
estimate for the error, is $\xi = 0.11 \pm 0.02$ fm.
We will come back to these results in
Section ~\ref{lambda} where we will compare them with those
obtained for the dual penetration length $\lambda$.

\subsection{Independence of the abelian projection}
\label{fixedgauge}

In the present Section we will discuss the possible dependence of
$\xi$ on the abelian projection chosen to define $\mu$.
The natural physical expectation is that $\xi$ be an universal
quantity characterizing the Yang-Mills vacuum, hence independent of the
particular abelian projection chosen.
This is consistent with 't Hooft ansatz that all abelian projections
are equivalent to each other: that equivalence also emerges
from numerical determinations of
$\langle \mu \rangle$, which have clearly showed that
$\pardis$ being zero or non zero is a gauge independent
statement~\cite{artsu2,artsu3,artran}.
A possible theoretical argument is the following: the operator
$\mu$ defined in one particular abelian projection creates a
magnetic charge in \hbox{every} other abelian projection~\cite{artfull2,abelind};
this implies
that the lowest mass state coupled to $\mu$ should be universal,
\ie $\xi$ should be independent of the abelian projection chosen.
In order to test that hypothesis we have repeated our measurements for
the abelian projection defined by diagonalizing
$P(\vec n, \hat t)$ on each lattice site, where $P(\vec n, \hat t)$ is
the Polyakov loop at the spatial site $\vec n$ starting at time $\hat t$.

The updating procedure in this case is not as simple 
as in the case of the random abelian projection:
changes in Polyakov loops modify the abelian projection and
as a consequence also the modified action $\tilde S$,
which therefore is not a linear function of the temporal links.
On those links usual heat-bath or
over-relaxation updatings are not possible and 
we have used a metropolis algorithm.
Numerical strategies for noise reduction like link integration
are no more feasible and as a consequence there is a considerable
increase in computational effort with respect to the case
of the random abelian projection.

\begin{figure}
\begin{center}
\includegraphics[width=8.5cm]{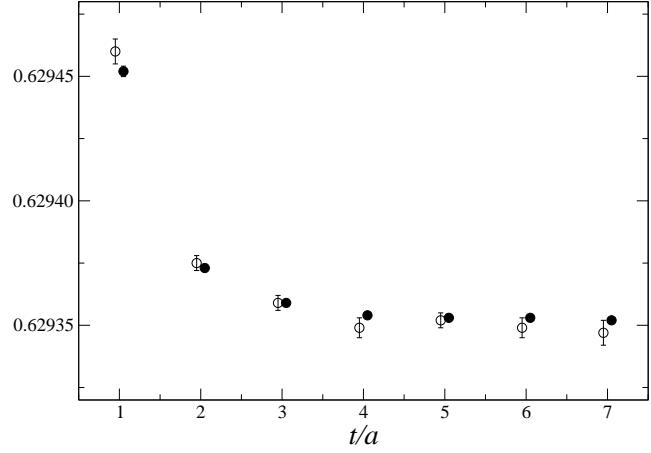}
\vspace{0.6cm}
\caption{Mean modified plaquette in the random gauge and
in the Polyakov gauge: an offset as been added in order to compare
 the two sets of data, which lead to compatible correlation lengths.
\label{xicmp}
}
\end{center}
\end{figure}

We have performed a numerical simulation in the 
Polyakov gauge at $\beta = 2.4$ on a $12^3\times20$ lattice. 
The results obtained in this case show a good 
agreement with the determination in the random gauge, as can 
be appreciated from Fig.~\ref{xicmp}. A fit according to Eq.~(\ref{fitfunc})
gives $\hat\xi(\beta=2.4)=1.3\pm0.8$, which is in agreement, even if
within the large errors, with the value obtained in the random gauge.

\subsection{Comparison with a different approach}
\label{rhotilde}

An alternative way to study the temporal correlator of
the disorder parameter, inspired by studies in
gaugeball spectroscopy~\cite{Majumdar}, has been introduced for 
the $U(1)$ pure gauge theory in Ref.~\cite{luca}, 
and consists in considering a new observable $\tilde \rho$, 
which is the derivative  
of $\ln \langle \bar{\mu}(\hat t,\vec{n})\mu(0,\vec{n}) \rangle$
with respect to the adimensional temporal distance $\hat t$ 
in place of the inverse gauge coupling $\beta$. The expected behaviour
for $\tilde \rho$ can be easily derived from Eq.~(\ref{betcorrel})
\beq
\label{eq:rhotilde}
\tilde \rho ( \hat{t} ) \equiv \frac{d}{d \hat t}
\ln \langle \bar{\mu}(\hat t,\vec{n})\mu(0,\vec{n}) \rangle \simeq
- \left(   
\hat M + \frac{3}{2 \hat t} 
\right)
\frac{\gamma e^{-\hat M \hat t} / \hat t^{3/2}}
     {\langle \mu \rangle^2 + \gamma \, {e^{- \hat M \hat t}}/~{{\hat
	   t}^{3/2}}} \; .
\eeq
Two features of the new observable $\tilde \rho$ are apparent from 
Eq.~(\ref{eq:rhotilde}):
1) the disconnected large distance contribution has
disappeared in taking the derivative, so that $\tilde \rho$ takes
into account only the interesting connected piece without any
noisy background; 2) there are
only 3 parameters ($\hat M$, $\gamma$ and $\pardis^2$) to be fitted.
Both things could contribute to make $\tilde \rho$ a better observable
than $\rho$ in order to extract the correlation length 
$\hat \xi = \hat M^{-1}$. To test that possibility and 
to have an independent check of the results presented in 
Section~\ref{rhoresults}, 
we have repeated our determination of $\xi$ by measuring the 
correlator $\tilde \rho (\hat t)$. 

\begin{figure}
\begin{center}
\includegraphics[width=9.5cm]{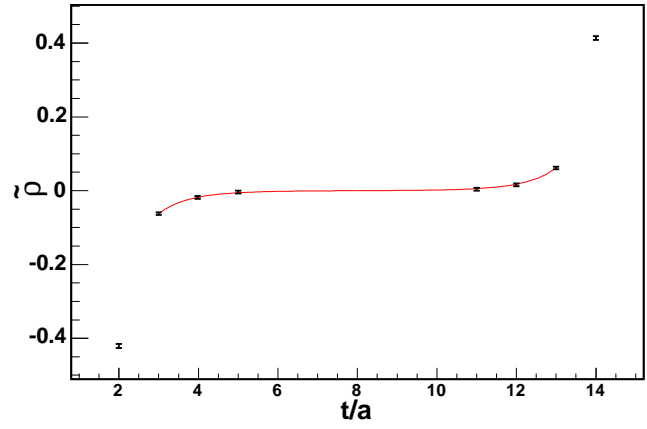}
\vspace{0.5cm}
\caption{Parameter $\tilde \rho$ measured at $\beta = 2.5115$ on a 
$12^3 \times 16$ lattice.
\label{fig:rhotilde}
}
\end{center}
\end{figure}

A drawback of $\tilde \rho$ is that its definition on the lattice
requires a prescription for the discretized derivative,
implying the possible presence of further systematic effects due to the
finite lattice spacing.
Let us rewrite 
the $\langle \bar \mu \mu \rangle$ correlator in the form
\beq
\langle \bar{\mu}(\hat t)\mu(0) \rangle = 
\frac{\int \left( {\cal D}U \right)  e^{-\beta {(S + \Delta S_0 +
      \Delta S_{\hat t})}}}{\int \left( {\cal D}U \right)  e^{-\beta {S}}} \; ,
\eeq
where $\Delta S_0$ and $\Delta S_{\hat t}$ indicate the changes
in the action in correspondence of the monopole creation and
destruction operators respectively; we will define $\tilde \rho$ by
taking the symmetric derivative, which can be written as 
(see Ref.~\cite{luca} for details):
\beq
\tilde \rho (\hat t) =
- {\beta \over 2} \, \left\langle ( \Delta S_{\hat t+1} - 
  \Delta S_{\hat t-1}  ) \right\rangle_{S + \Delta S_0 +
      \Delta S_{\hat t}} \; 
\label{obsrhotilde}
\eeq
We have performed our measurement at $\beta = 2.5115$ on a lattice 
$12^3 \times 16$. Our results are reported in Fig.~\ref{fig:rhotilde}
together with the best fit result obtained using Eq.~(\ref{eq:rhotilde})
after taking into account the periodic boundary conditions in
the time direction\footnote{Since $\tilde \rho (\hat t)$ is the first
derivative of the temporal correlator, it is an odd function with respect
to $\hat t = N_t/2$, as is clearly verified from
Fig.~\ref{fig:rhotilde}.}: the first point included in the fit has been 
$\hat t_0 = 3$. We have obtained
$\hat \xi = 1.32(25)$ ($\xi = 0.110(21)$ fm in physical units) with
$\chi^2/\mbox{d.o.f.} = 1.8/3$. 
We conclude that the agreement with the results obtained by 
measuring $\rho$ is very good (see Tables~\ref{xitable} and 
\ref{xiscalingtab}): this consistency gives an indication
that systematic effects in the determination of $\xi$ 
are under control both when $\rho$ or $\tilde \rho$ are used as observables.
The precision on $\xi$ is similar to that obtained in 
Section~\ref{rhoresults}: since a comparable statistics has been used,
we conclude that the benefit of dealing with a connected observable
is not very significant.

\section{Determination of $\lambda$ and typology of the vacuum}
\label{lambda}

Several consistent determinations of the parameter $\lambda$ can be found in the 
literature~\cite{suz94,cea95,bali98,suz99,suz03,hay05}: in this 
Section we will present an independent one obtained from the study
of the flux tube profile between two static color charges.

We have analyzed, for two different values of the inverse gauge
coupling, $\beta = 2.5115$ and $\beta = 2.6$, the abelian projected
flux tube formed between a quark and an antiquark placed at $16$
lattice spacings apart from each other (corresponding
respectively to 1.33 and 0.99 fm); the Maximal Abelian gauge has been
chosen to define the abelian projection. In particular we have
studied the correlation of the plaquette operator with a Wilson loop
$W(R,T)$ with $R=16$ and $T=6$, on a lattice
$24\times24\times32\times24$, with the longer dimension along
the $q \bar q$ axis. We adopt the following prescription for the
electric field~\cite{hay05}:
\begin{equation}
E_i=\frac{\langle\mbox{tr} \left(W^{AbPr}(R,T)\prod{}^{AbPr}_{0i} \right)\rangle
} {\langle\mbox{tr}\left(W^{AbPr}(R,T)\right)\rangle} - \frac{\langle\mbox{tr}\left(
W^{AbPr}(R,T)\right) \mbox{tr} \prod{}^{AbPr}_{0i}\rangle}
{2\,\langle\mbox{tr}\left(W^{AbPr}(R,T)\right)\rangle} \label{ourEi} \, ;
\end{equation}
which is the definition satisfying the
Maxwell equations on the lattice~\cite{hay05}. 
The flux tube profile has been studied at half a way
between the two charges and at an equal temporal distance between
the creation and annihilation times of the $q \bar q$ pair.
The Jackknife method for correlated
quantities~\cite{jack} has been used in the statistical analysis.

Having chosen a quite long flux tube, noise reduction is a critical point
of our computation: we have adopted a standard cooling procedure,
looking for a stable plateau in $\lambda$ as a function
of the cooling steps performed.

We have fitted our data according to 
the solution $E_z=A K_0(\frac{d}{\hat\lambda})$ of London equation $\nabla^2 E_z =\frac{1}{\lambda^2} E_z$: 
since that is expected to be valid beyond a certain distance $d$ from the 
flux tube axis, where the effects of the non superconductive core
are absent, 
we look for a plateau of $\hat\lambda$ with respect to the minimum
distance $d_0$ included in the fit.

\begin{figure}[htbp]
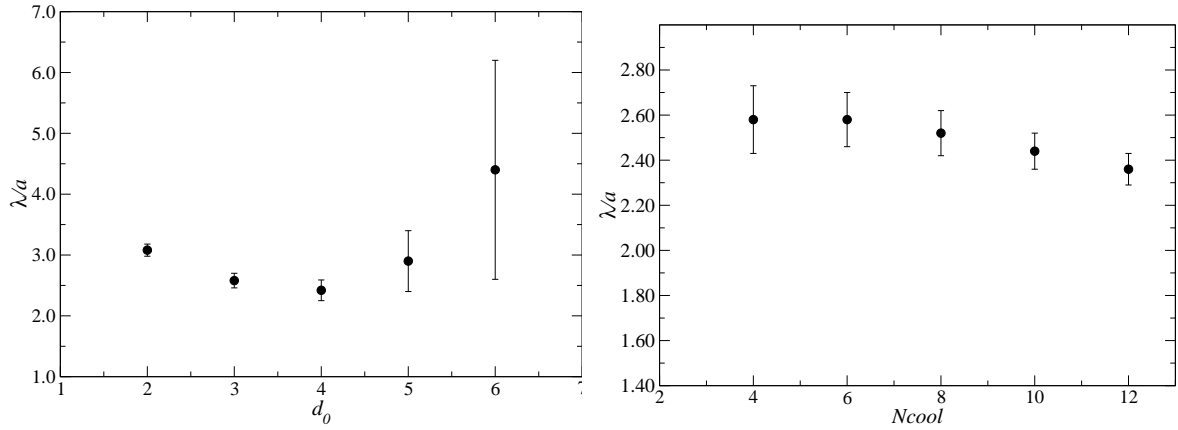
 \begin{center}
\begin{tabular}{cc}
\includegraphics[width=7.7cm]{lambdafitdepcool6.eps} & \includegraphics[width=7.7cm]{lambdacooldepd3.eps}\\
\end{tabular}
\vspace{0.5cm}
\caption{$\hat \lambda$ as a function of the fit
starting point $d$ (left) and as a function of the number of cooling
steps (right) at $\beta=2.6$. \label{lambdacoolfitdep}}
\end{center}
\end{figure}

In figure~\ref{lambdacoolfitdep} we report the dependence of $\hat
\lambda$ measured at $\beta =  2.6$ both as a function of the
fit starting point $d_0$ at a fixed number of cooling steps
($N_{cool}=6$) and as a function of the number of cooling steps at a
fixed fit starting point ($d_0=3$). A plateau is visible
in both cases and we choose $d_0 = 3$ and 6 cooling steps as a
reference.

\begin{figure}
\begin{tabular}{ll}
\includegraphics[width=8.3cm]{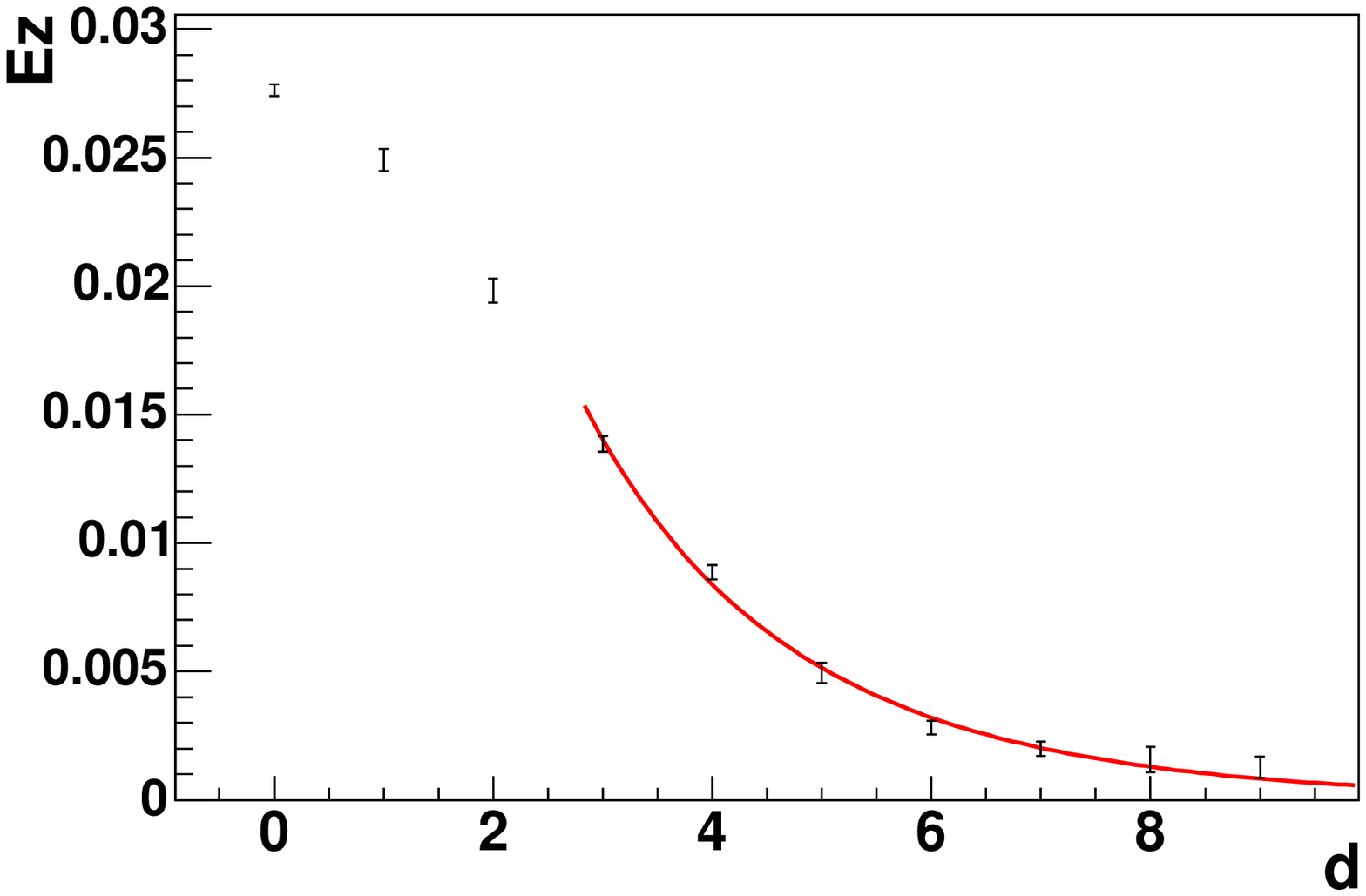} &
\includegraphics[width=8.3cm]{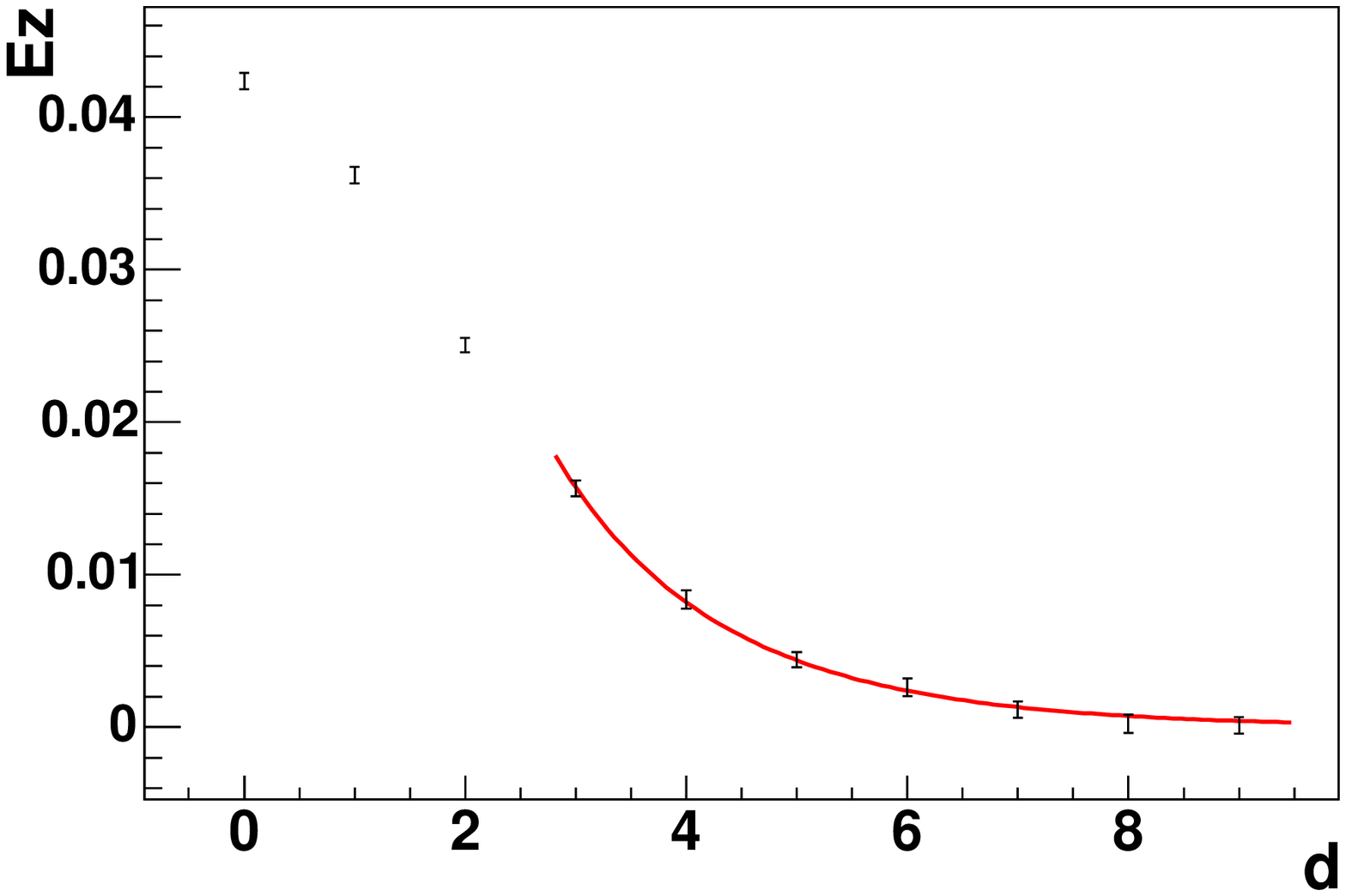}
\end{tabular}
\vspace{8pt} \caption{Profile of $E_z$ after $6$ 
cooling steps for a $24\times24\times32\times24$ lattice at
$\beta=2.6$ (left) and at $\beta=2.5115$ (right). The black line
refers to a fit with the function
$A K_0(\frac{d}{\hat \lambda})$.
\label{Ez(d)}}
\end{figure}

Our fitted value at $\beta=2.6$ (Fig.~\ref{Ez(d)}, left) is $\hat\lambda(\beta=2.6)=2.58\pm0.12$;
a similar analysis at $\beta=2.5115$
(Fig.~\ref{Ez(d)}, right) leads to $\hat\lambda(\beta=2.5115)=1.96\pm0.08$.
Converting our results into physical units (see
 Section~\ref{rhoresults})
 we obtain $\lambda=0.163\pm0.007$ fm at $\beta=2.5115$ and
$\lambda=0.160\pm0.007$ fm at $\beta=2.6$ lattice in good agreement with
previous literature (as one of the latest determinations
we report $\lambda = 0.157 \pm 0.003$ from Ref.~\cite{hay05}).

\begin{figure}
\begin{center}
\includegraphics[width=9.6cm]{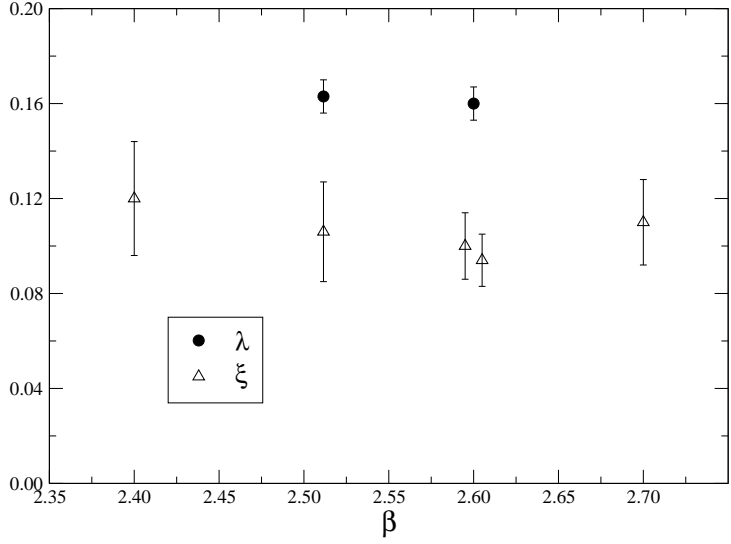}
\vspace{0.2cm}
\caption{$\xi$ and $\lambda$ 
in fermi units for different values of $\beta$. 
The two different values shown at $\beta = 2.6$, which have
been slightly split apart for the sake of clarity, correspond to 
two different monopole charges, $Q = 8$ and $Q = 2$ respectively.
\label{scaling}}
\end{center}
\end{figure}

In Fig.~\ref{scaling} we report
a summary of the results obtained for $\lambda$ and $\xi$ at the
different values of the lattice spacing.
While apparently $\xi$ is consistently lower than $\lambda$, it is
anyway clear that the two quantities are comparable, in agreement
with the findings of previous
literature~\cite{bali98,suz99,hay05,suz05}. Our conclusion is
therefore that the vacuum type of pure gauge QCD with two colors is
close to the type I~-~type II boundary, even if marginally of type
II.

Our result can be further clarified by looking
for observable consequences of the QCD vacuum being a type I or type II
superconductor: that will be the subject of next Section.

\section{Analysis of flux tubes interactions}
\label{2tube} Another direct (but qualitative) method to identify
the typology of the vacuum is to look at the behavior of two close parallel
flux tubes: in a type II superconductor nearby flux tubes
repel each other, while attraction is expected for a type I superconductor. 

We have looked at the flux tubes of two $q \bar q$ pairs placed with their axes parallel
to each other (and along the $z$ direction).
We have performed two simulations at $\beta = 2.6$, placing 
each quark at a distance of $16$ lattice spacings from the respective
antiquark and considering two different distances $D$ between the 
two parallel $q \bar q$ axes, $D = 4$ and $D = 5$;
the lattice chosen is again a  
$24\times24\times32\times24$ and following Section~\ref{lambda}, we have determined the abelian
electric field in presence of two parallel Wilson loops 
$W_1(R,T)$ and $W_2(R,T)$ at distance $D = 4\, a$ or 
$D = 5\, a$, with $R = 16$ and $T = 6$.
The quantity we look at is

\beq
E_i=\frac{\langle \mbox{tr} \left(W_1^{AbPr}(R,T)
W_2^{AbPr}(R,T)\prod{}^{AbPr}_{0i} \right)\rangle }
{\langle\mbox{tr}\left(W_1^{AbPr}(R,T)W_2^{AbPr}(R,T)\right)\rangle} -
\frac{\langle\mbox{tr}\left( W_1^{AbPr}(R,T) W_2^{AbPr}(R,T) \right)
\mbox{tr} \prod{}^{AbPr}_{0i}\rangle}
{2\,\langle\mbox{tr}\left(W_1^{AbPr}(R,T)W_2^{AbPr}(R,T)\right)\rangle}
\label{doubledipole}
\eeq
that is the generalization of 
Eq.~(\ref{ourEi}) satisfying the 
Maxwell equations on the lattice. Our statistics consist of
about $6 \cdot 10^4$ decorrelated configurations for each simulation.

\begin{figure}[!htbp]
\begin{center}
\begin{tabular}{ll}
\includegraphics[width=8.9cm]{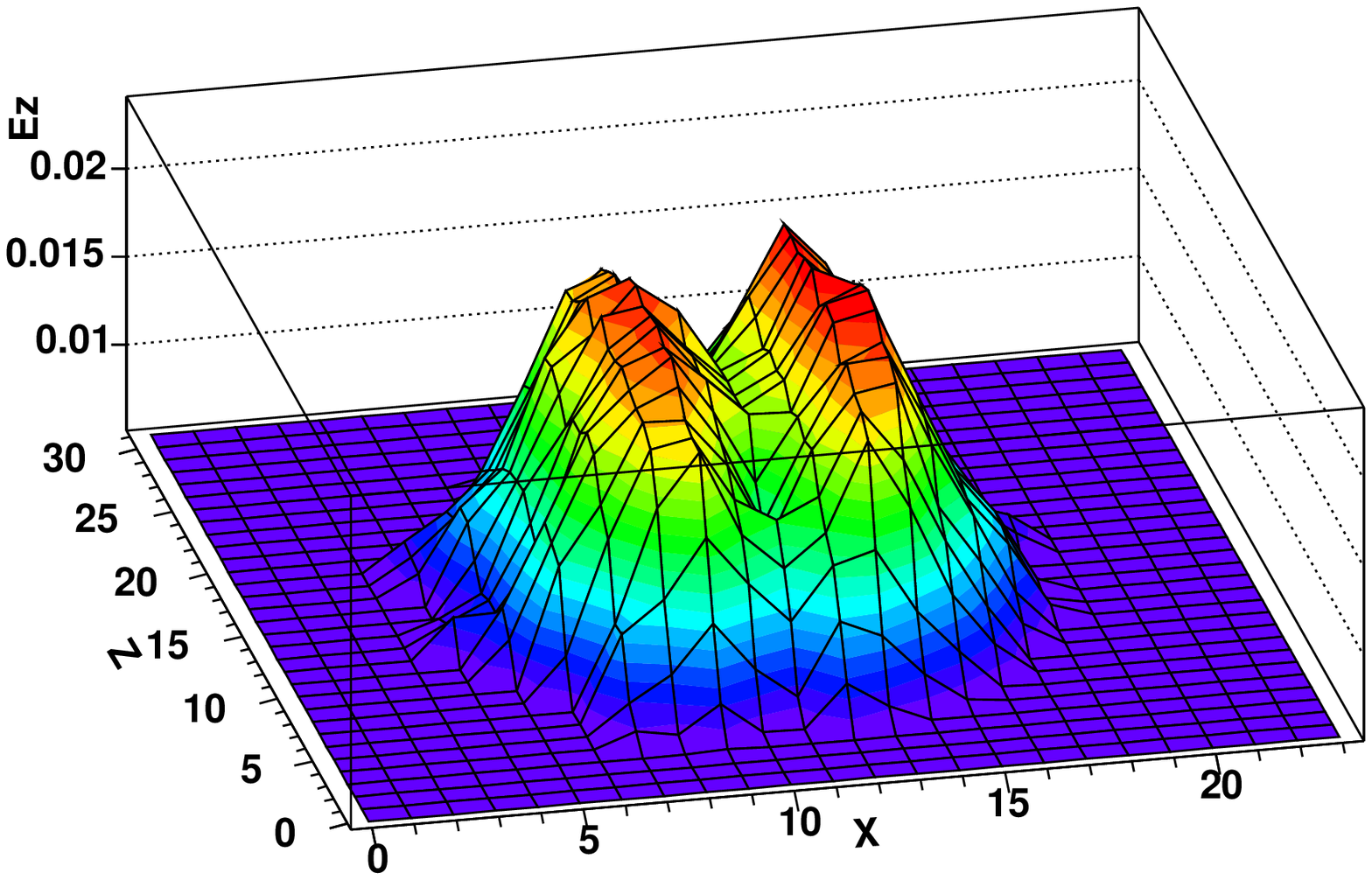} &
\includegraphics[width=8.3cm]{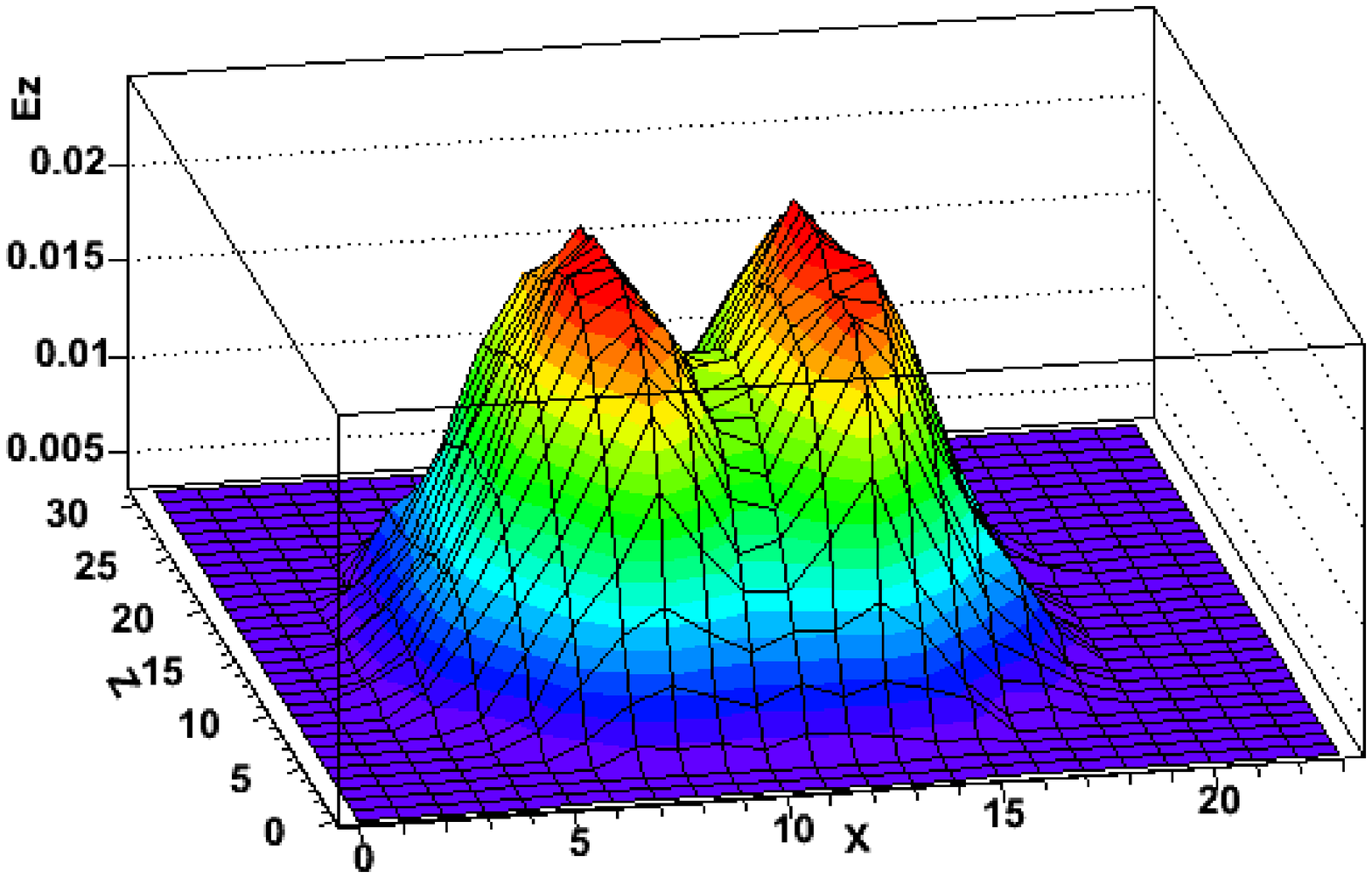} 
\end{tabular}
\caption{Profile of the two interacting flux tubes on
the $xz$ plane placed at different relative distances ($D = 4$ and $D = 5$): 
the two $q \bar q$ axes are placed respectively at
$x = 9$ and $x = 13$ (left), and at $x = 8$ and $x = 13$ (right). 
The electric field has been measured after $6$ cooling steps.
\label{twostrings}}
\end{center}
\end{figure}

We focus on the longitudinal $E_z$ component in the $xz$ plane reported in
Fig.~\ref{twostrings} where errors on $E_z$
are of the order of $5\%$ for almost all points. As in Section~\ref{lambda} we have
used cooling for noise reduction: all data showed have been obtained
after 6 cooling steps.

No evident repulsive or attractive behaviour can be appreciated
from Fig.~\ref{twostrings}, however we have tried 
a quantitative analysis of flux tube deflection by measuring the average distance between the two tubes
and comparing it with the distance $D$ between the two $q \bar q$ axes.
The average distance has been taken over the central part of the flux
tubes, including 9 lattice sites for each tube.
We have defined the position of the flux tube in two different ways:
first by the position of the local maximum 
for $E_z$, 
secondly by the weighted average
position over the three lattice sites closest to the $q \bar q$
axis, using $E_z$ as a weight. We call the two definitions $d_M$ and $d_W$ respectively.
 
In Table~\ref{deflection} we report the data obtained for the 
deflections ($d_M - D$) and  ($d_W - D$) as a function of $D$: 
a positive/negative value corresponds to a repulsive/attractive
behaviour. While deflections are nearly compatible with zero 
at $D = 5$, some signal appears when the 
flux tubes are closer to each other, at $D = 4$. 
Although we have no clear sign of flux tube repulsion, we consider this as an important hint in that 
direction, even more if we consider that the superposition of the two 
flux tubes in the central region should bias our result in the
opposite direction, leading to negative values of ($d_M - D$) and ($d_W - D$).

In Section~\ref{lambda} we concluded that the vacuum is close to the 
type I - type II boundary, even if marginally on the type II side:
that would imply a weak repulsive interaction between parallel flux
tubes. That is consistent with the result of the 
present Section, \ie that there are signs of weak repulsive interaction 
as the distance between the two flux tubes is decreased. The still large
uncertainties as well as the finite lattice spacing place a limit
on the observable flux tube deflection.
While the aim of the study presented in this Section was only to look
for a possible evident signal of flux tube deformation, more refined investigations can be done, including a detailed
analysis of Wilson loop interactions and a quantitative comparison with the determinations of $\xi$ and 
$\lambda$, after also taking properly into account the quantum fluctuations of the flux tube. 
 That is beyond the purpose of the present
study and will be the subject of future investigations.

\section{Conclusions}
\label{final}

The aim of our study was that of understanding which type of dual
superconductor is realized in the vacuum of $SU(2)$ pure gauge
theory. 
We have followed two different strategies: a numerical determination of
the parameters $\xi$ and $\lambda$ and an analysis of the
interaction between parallel flux tubes.

To determine $\xi$ we have studied
the temporal correlator of an operator $\mu$
which creates a magnetic monopole and whose vacuum expectation value
has been shown to be an order parameter for dual superconductivity
both in quenched and in full QCD.
The greatest difficulty in our measurement derives from the
necessity of isolating an exponentially decaying signal from a large
background: that is why a
very high precision is needed and numerical strategies such as analytic
link integration have been used. We have determined $\xi$ for four 
different values of the inverse coupling $\beta$ and, in some cases,
for different lattice volumes and for different charges $Q$ of the
magnetic monopole. We have explicitely 
checked that our results are independent of the abelian projection used 
to define $\mu$, that they have no significant dependence on the
monopole charge $Q$ and that they are not affected by finite volume
effects within our statistical uncertainties.
Data are compatible within errors with the correct scaling to the
continuum limit, even if a reliable extrapolation to that limit cannot 
still be performed. 
Results are summarized in Table~\ref{xiscalingtab} and our estimate for the 
correlation length, based on the determination at the largest value of
$\beta$, is $\xi = 0.11 \pm 0.02$ fm.
 We have also repeated our measurement
using an alternative way to study the temporal correlator~\cite{luca},
obtaining a good agreement. We would like to stress  
the consistency of our results with those reported in
Ref.~\cite{suz05}, which were obtained through a completely different method,
consisting in the study of the correlations of monopole currents
around the flux tube.

To determine $\lambda$
we employed the usual analysis of the longitudinal component of the
abelian chromoelectric field inside the flux tube. We performed our
measurement at two different couplings,
$\beta = 2.5115$ ($\lambda = 0.163(7)$ fm) and $\beta = 2.6$
($\lambda = 0.160(7)$ fm), 
obtaining a  good agreement with previous literature.

Our determinations show that $\xi$ is smaller than $\lambda$, even if 
comparable to it in magnitude: this indicates that the vacuum of pure gauge 
QCD with two colors is close to the type I~-~type II boundary and
marginally of type II. This is consistent with our direct
investigation presented in Section~\ref{2tube}, 
showing some weak signals of repulsive interactions as two parallel
flux tubes are brought closer to each other.

One way to improve our results would be to make a more precise determination
of $\xi$: the great noise  in the signal obtained for the
temporal correlator has been a limitation and no significant
improvement has been achieved when adopting the alternative approach 
proposed in Ref.~\cite{luca}.
One reason for the problems encountered can be traced back 
to the small value of $\xi$ itself
($\xi \sim 0.1$ fm), which makes the signal to fade away very
rapidly after a few lattice spacings.
One possible strategy to overcome this limitation
could be to make use of anisotropic lattices with a 
very small lattice spacing in the temporal direction; we will also consider
direct determinations 
of the correlator 
$\langle \bar{\mu}(t,\vec{x})\mu(0,\vec{x}) \rangle$, in place
of its derivatives $\rho (t)$ or $\tilde \rho (t)$, using a technique
recently developed for the $U(1)$ gauge theory~\cite{mu_u1}.
Finally, a more refined 
investigation of flux 
tube interactions could be performed, using also finer
lattice spacings in order to be 
more sensitive to small flux tube deflections.

As an extension of our study, we will repeat in the future
the determination of $\xi$ also around or slightly above the
deconfinement critical temperature $T_c$, where $\xi$ could be directly 
related to the mass of physical magnetic monopoles: that could be
quite relevant from a phenomenological point of view, if the
hypothesis~\cite{zak-che} of light monopole degrees of freedom
populating the quark gluon plasma slightly above the transition is correct.
Of course it would be also of fundamental importance to extend 
our investigation to the case of pure gauge theory with 3 colors
and eventually to full QCD.

\section*{Acknowledgements}

\vskip 5mm

It is a pleasure to thank A. Di Giacomo 
for useful comments and discussions. Numerical simulations have
been run on two PC farms at INFN - Genova and at CNAF - Bologna.
This work has been partially supported by MIUR.

\begin{table}[htp]
\begin{center}
\begin{tabular}{|c|c|c|c|c|c|}
\hline
                                                 &     $\hat t_0=1$    &   $\hat t_0=2$   &   $\hat t_0=3$    &   $ \hat t_0=4$  & $\hat t_0=5$ \\
\hline
$12^3\times16$, $\beta=2.4$, $Q=2$        &   $0.857\pm0.021$   &  $ 1.07\pm0.07$    & $ 1.02\pm0.20$     &    $4.5\pm3.8$ & \\
\hline
$12^3\times20$, $\beta=2.5115$, $Q=2$     &                     &   $1.28\pm0.08$    &  $1.36\pm0.21$       &  $1.8\pm1.5$          & \\
\hline
$16^3\times20$, $\beta=2.5115$, $Q=2$     &    $0.99\pm0.03$    &   $1.14\pm0.10$    &   $1.28\pm0.25$    &    $1.0\pm0.4$ & \\
\hline
$20^3\times20$, $\beta=2.6$, $Q=2$        &          $1.24\pm0.03$    &   $1.33\pm0.05$    &   $1.52\pm0.17$    &   $3.6\pm1.5$  & \\
\hline
$20^3\times20$, $\beta=2.6$, $Q=8$        &   $1.06\pm0.03$                   &   $1.32\pm0.08$    &   $1.62\pm0.23$   &   $1.8\pm0.6$   &  $1.2\pm1.0$\\
\hline
$20^3\times20$, $\beta=2.7$, $Q=8$        &    $1.34\pm0.02$      &    $1.51\pm0.04$   &   $1.78\pm0.11$   &   $2.4\pm0.4$   &  $1.7\pm0.8$\\                  
\hline
\end{tabular}
\caption{Dependence of $\hat \xi$ on the fit starting
point $\hat t_0$ for our sets of numerical simulations 
(fit according to Eq.~\ref{fitfunc}). \label{xitable}
}
\end{center}
\end{table}

\begin{table}
\begin{tabular}{|c|c|c|}
\hline
$\beta$  &   $a(\beta)$ & $\xi$   \\
\hline
$2.4$     &    $0.118$ fm & $0.126\pm0.008$ fm \\
\hline
$2.5115$  &    $0.083$ fm & $0.106\pm0.021$ fm \\
\hline
$2.6$     &    $0.062$ fm & $0.094\pm0.011$ fm ($Q=2$),\ \ \ \ $0.100\pm0.014$ fm ($Q=8$)\\
\hline
$2.7$     &    $0.046$ fm & $0.110\pm0.018$ fm \\
\end{tabular}
\vspace{0.3cm}
\caption{Lattice spacing $a$ and correlation length 
$\xi$ for different values of $\beta$. \label{xiscalingtab}}
\end{table}

\begin{table}
\begin{tabular}{|c|c|c|}
\hline
$\,\,\,\,\,\,D\,\,\,\,\,\,$  &   $d_M - D$ & $d_W - D$   \\
\hline
4     &  0.44(17) &  0.017(9)  \\
\hline
5  &  0.22(14) &  -0.007(15) \\
\hline
\end{tabular}
\vspace{0.3cm}
\caption{Average deflection (in lattice spacing units) at 
$\beta = 2.6$ for two parallel flux tubes as
a function of the distance $D$ between the $q \bar q$ axes.
See text for the definition of $d_M$ and $d_W$.
\label{deflection}}
\end{table}

\newpage

\vspace{1.0cm}

\newpage

\vspace{1.6cm}

\newpage

\vspace{1cm}

\end{document}